\documentclass[12pt]{article}
\usepackage{natbib} 

\usepackage[latin1]{inputenc}
\usepackage{amsthm}
\usepackage{amsmath}
\usepackage{amsfonts}
\usepackage{amstext}
\usepackage{algorithm}
\usepackage[noend]{algpseudocode}
\usepackage{graphicx}
\usepackage{natbib}
\usepackage{subcaption}
\usepackage[dvipsnames]{xcolor}
\usepackage{soul}

\usepackage{setspace}

\newtheorem{definition}{Definition}
\newtheorem{proposition}{Proposition}
\newtheorem{corollary}{Corollary}
\newtheorem{example}{Example}[section]

\newcommand\numberthis{\addtocounter{equation}{1}\tag{\theequation}}

\begin{document}
		\title{Exploiting the Superposition Property of Wireless Communication for Max-Consensus Problems in Multi-Agent Systems}
		\author{
			Fabio Molinari\thanks{F. Molinari is with the Control Systems Group - Technische Universit\"at Berlin, Germany.}
			\and S\l awomir Sta\'nczak\thanks{S. Sta\'nczak is with the Network Information Theory Group - Technische Universit\"at Berlin, Germany \& Fraunhofer Heinrich Hertz Institute, Germany.}			
			\and J\"org Raisch\thanks{J. Raisch is with the Control Systems Group - Technische Universit\"at Berlin, Germany \& Max-Planck-Institut f\"ur Dynamik komplexer technischer Systeme, Germany.}
		}
			
		\maketitle
		\begin{abstract}
			\footnote{This work was funded by the German Research Foundation (DFG) within their priority programme SPP 1914 "Cyber-Physical 
				Networking (CPN)".	
			}
			This paper presents a consensus protocol that achieves max-consensus in multi-agent systems over wireless channels. 
			Interference, a feature of the wireless channel,
			is exploited:
			each agent receives a superposition of broadcast data, rather than individual values.
			With this information, the system endowed with the proposed consensus protocol
			reaches max-consensus in a finite number of steps.
			A comparison with traditional approaches shows that the proposed consensus protocol achieves a faster convergence.
		\end{abstract}
	\section{Introduction}	
	Consensus is a useful concept
	in cases where several agents need to achieve agreement over a variable of common interest, \cite{ren2007information}.
	Each agent retains a local guess of this variable, which is referred to as its \textit{information state}.
	The ingredients of a consensus strategy are, first, an exchange of information between agents (\textit{communication}), 
	then, an update of information states according to a suitable algorithm (\textit{computation}). 
	
	This paper focuses on a specific consensus called max-consensus (\cite{olfati2004consensus}).
	The convergence of max-consensus algorithms has been studied, e.g., via max-plus algebra techniques  
	in \cite{nejad2009max}, showing that an agreement is achieved 
	in a finite number of steps.
	
	Convergence rate is an important property.
	Many applications, as \cite{molinari2018automation}, show that getting to an agreement as fast as possible is a main issue,
	also for safety reasons. 
	
	Traditionally, communication and computation have been treated as two distinct aspects.
	However, \cite{goldenbaum2013harnessing} claim that, in a wireless communication framework, a considerable increase of convergence rate can be achieved by merging these two aspects.
	Following this path, \cite{molinari2018exploiting} propose an average-consensus protocol which harnesses the interference of the wireless channel,
	thus achieving a much faster agreement.
	In the proposed scheme, all agents broadcast their values simultaneously, instead of creating orthogonal channels. 
	Via an appropriate consensus protocol,
	the received interfered signals can lead to an agreement.	
	The broadcast property of the wireless channel has been harnessed for achieving
	max-consensus via randomized protocols in \cite{iutzeler2012analysis}.
	To the best of our knowledge, no deterministic broadcast protocols for max-consensus have been proposed so far.
	
	In the remainder of the paper, a general consensus problem over wireless channel is described in Section \ref{sec:problemdesc}. 
	A broadcast max-consensus protocol is presented in Section \ref{sec:asympt} and proven to converge asymptotically to an agreement.
	Achieving consensus in a finite number of steps is possible by using the consensus algorithm presented in Section \ref{section:finite}.
	Simulations are then analyzed in Section \ref{sec:simul}. Finally, in Section \ref{sec:concl}, concluding remarks are provided.


		\subsection{Notation}
		{We use $\mathbb{N}$, $\mathbb{R}$, $\mathbb{R}_{>0}$, and $\mathbb{R}_{\geq0}$ to denote the set of positive integers, the set of real numbers, the set of positive real numbers, and the set of nonnegative real numbers, respectively.	}	
		The vector $\mathbf{1}_n\in\mathbb{R}^n$ denotes the all-ones vector of dimension $n\in\mathbb{N}$.
		The transpose of a matrix $A\in\mathbb{R}^{n\times m}$ is denoted by $A'$.
		The cardinality of a {finite} set $\mathcal{S}$ is {denoted} by $|\mathcal{S}|$.
		The indicator function $I_A(x):X\rightarrow\{0,1\}$ on a set $X$ is defined to be
		$$
		I_A(x):=
		\begin{cases}
			1\ \ \ \text{if } x\in A,\\
			0\ \ \ \text{else,}
		\end{cases}
		\ A\subseteq X.
		$$
		{An undirected} graph is a pair $(\mathcal{N},\mathcal{A})$, where $\mathcal{N}$ is a set of nodes, while $\mathcal{A}$ is the corresponding set of arcs.
		If an arc connects nodes $i,j\in\mathcal{N}$, then $(i,j)\in\mathcal{A}$.
		Given a graph, we define a \textit{path} to be a sequence of nodes, in which each adjacent pair is connected by an arc.
		A graph is said to be \textit{connected} if there exists a path between any distinct pair of nodes.
		Given a node $i\in\mathcal{N}$, its set of neighbors is denoted by $N_i$ and {we have}
		$$N_i=\{ j\in\mathcal{N}\mid (i,j)\in\mathcal{A} \} .$$
		
	\section{Problem Description}
	\label{sec:problemdesc}
	{We consider a discrete-time multi-agent system with $n\geq1$ agents communicating over a wireless network modeled by the graph 
	$(\mathcal{N},\mathcal{A})$ where $\mathcal{N}=\{1,\dots,n\}$. In what follows, let $k\in\mathbb{N}$ be the time index.}
	{The system has a variable of common interest which the agents have to agree on.
	Each agent $i\in\mathcal{N}$ has its initial estimate of this variable}, i.e., $x_{i_0}\in\mathbb{R}_{\geq0}$, which we refer to as its initial \textit{information state}. 
	The objective of the consensus protocol
	is to enable all agents to reach an agreement over their information states. 
	To this end, 
	each agent, say agent $i$, will dynamically update its information state, i.e. $x_i:\mathbb{N}\rightarrow\mathbb{R}_{\geq0}$, according to the consensus protocol and to the information received from its neighbouring agents. Let $\forall i\in\mathcal{N}$, $\mathbf{x}_{N_i}:\mathbb{N}\rightarrow\mathbb{R}_{\geq0}^{|N_i|+1}$ be a vector containing the information states of agents in the set $N_i\cup\{i\}$ at instant $k\in\mathbb{N}$, i.e. 	
	\begin{equation}
		\mathbf{x}_{N_i}(k)=[x_i(k), x_{j1}(k),\dots,x_{j_{m_i}}(k)]',
	\end{equation} 
	where $j_1,\dots,j_{m_{i}}\in N_i$ and $m_i=|N_i|$.
	Accordingly, a general discrete-time consensus protocol is
	\begin{equation}
		\label{eq:generalConsProt}
		x_i(k+1)=f_i(\mathbf{x}_{N_i}(k)),
	\end{equation}
	where $f_i:\mathbb{R}_{\geq0}^{|N_i|+1}\rightarrow\mathbb{R}_{\geq0}$ 
	and, $\forall i\in\mathcal{N},\ x_i(1)=x_{i_0}$.
	Let $\mathbf{x}(k)$ be the vector of all information states, i.e. $\forall i\in\mathcal{N},\ \left[\mathbf{x}(k)\right]_i=x_i(k)$.
	Then, we say that the multi-agent system converges to max-consensus if
	\begin{equation}
		\forall i\in\mathcal{N},\ \lim\limits_{k\rightarrow\infty} x_i(k) = x^* = \max(\mathbf{x}(0)).
	\end{equation}
	Moreover, if the system converges to max-consensus in a finite number of steps, then {the strategy is referred to as}
	\textit{finite-time max-consensus}. This is formally achieved if $\exists \bar{k}\in\mathbb{N}$, such that $\forall k>\bar{k}$, 
	\begin{equation}
		\forall i\in\mathcal{N},\ x_i({k})= x^* = \max(\mathbf{x}(0)).
	\end{equation}	
	\begin{definition}
		Given $k\in\mathbb{N}$, an agent $i\in\mathcal{N}$  {is said to be maximal at $k\in\mathbb{N}$ if} $x_i(k)=\max(\mathbf{x}(k))$.
	\end{definition}
	
		\subsection{Traditional Max-Consensus Protocols}
		{Widely-considered max-consensus protocols are of the form}
		\begin{equation}
			\label{eq:maxcons}
			x_i(k+1)=\max(\mathbf{x}_{N_i}(k)).
		\end{equation}
		\cite{nejad2009max}, under the assumption of a time-invariant and connected network topology, show that the consensus protocol (\ref{eq:maxcons}) ensures max-consensus in a finite-number of steps.
		The wireless communication is usually based on orthogonal channel access methods, which establish interference-free transmission links between neighbours and provide each agent with the knowledge of individual information states of the neighbors.
		
		However, {by the data processing inequality} \cite[p. 32]{cover2012elements}, the amount of information contained in $\max(\mathbf{x}_{N_i}(k))$ is in general less than the amount carried by the vector $\mathbf{x}_{N_i}(k)$ itself. Therefore, reconstructing neighbors' information states appears a suboptimal strategy if the goal is to reconstruct only the maximum value of the information states. 
		\subsection{Interference model}
		Let $N_i=\{{j_1},\dots,{j_{m_i}}\}\in\mathcal{N}$ be a set of agents {that} transmit their respective information states to agent $i\in\mathcal{N}$ at discrete-time step $k\in\mathbb{N}$. 
		If orthogonal channel access methods are not used, then interference occurs.
		Accordingly, agent $i\in\mathcal{N}$ receives the signal $\zeta_i(k)$ which is a superposition of $x_{j_1}(k),\dots,x_{j_{m_i}}(k)$.
		This is often modeled by an affine model of the wireless multiple access channel (MAC), as in \cite{molinari2018exploiting}, i.e.
		\begin{equation}
			\label{eq:MAC}
			\zeta_i(k)
			=
			\sum_{j\in N_i} h_{ij}(k)x_j(k) + v_i(k),
		\end{equation}
		where 
		$\forall k\in\mathbb{N},\ h_{ij}(k)\in(0,1]\subset\mathbb{R}$ 
		are referred to as channel coefficients, and $\forall i\in\mathcal{N},\ \forall k\in\mathbb{N},\ v_i(k)$ is the receiver noise.
		In the following, we ignore both channel coefficients and receiver noise, as in \cite{goldenbaum2012nomographic}; accordingly, (\ref{eq:MAC}) reduces to an ideal MAC
		\begin{equation}
		\label{eq:MAC_ideal}
		\zeta_i(k)
		=
		\sum_{j\in N_i} x_j(k).
		\end{equation}
		{This shows that the nature of interference is \textit{superposition}}.
		
		\subsection{Nomographic Representation}
		By the \textit{superposition theorem} of \cite{kolmogorov1963representation}, any multivariate function $f:\mathbb{R}^n\rightarrow\mathbb{R}$ has a \textit{nomographic representation}
		\begin{equation}
			\label{eq:nomRepr}
			f_i(x_1,\dots,x_n)=\psi_i(\sum_{j=1}^n \phi_j(x_j)),
		\end{equation}
		where the univariate functions $\psi_i:\mathbb{R}\rightarrow\mathbb{R}$ and $\phi_j:\mathbb{R}\rightarrow\mathbb{R}$, $j=1\dots n$, are referred to as post-processing function and pre-processing functions, respectively. 
		
		By (\ref{eq:nomRepr}), and according to the interference model in (\ref{eq:MAC_ideal}), each function can be computed over the wireless channel by harnessing its interference property. 
		In the context of a consensus problem, 
		if each agent $i\in\mathcal{N}$ has to compute (\ref{eq:generalConsProt}),
		a procedure, which aims to merge communication and computation over the channel, is presented in Algorithm \ref{algo:nomo_algo} and has to be run $\forall k\in\mathbb{N}$.
		\begin{algorithm}
			\caption{}
			\label{algo:nomo_algo}
			\begin{algorithmic}
				\State $\forall j\in\mathcal{N}$, agent $j$ broadcasts the pre-processed value of its current information state, $\phi_j(x_j(k))$;
				\State $\forall i\in\mathcal{N}$, agent $i$ receives the superposed signals from neighbors, $z_i(k)=\sum_{j\in N_i} \phi_j(x_j(k))$;
				\State $\forall i\in\mathcal{N}$, agent $i$ computes $\psi_i(\phi_i(x_i(k))+z_i(k))$, thus getting the desired $f_i(\mathbf{x}_{N_i}(k))$.	
			\end{algorithmic}
		\end{algorithm} 		
		Computing a function over the wireless channel, by using its nomographic representation and the interference, results in a much faster and more efficient solution. A quantitative analysis can be found in \cite{goldenbaum2013harnessing}.
		
		\subsection{{Approximated Nomographic Representation}}
		\label{subsec:approx_nom_repr}
		However, \cite{buck1982nomographic} states that, in general,
		functions do not have a continuous real-valued nomographic representation. 
		In \cite{limmer2015simple} the idea of a nomographic approximation is suggested.
		However, as in the suggested nomographic approximations for the max-function the error is always positive,
		it accumulates over time in a consensus algorithm and will therefore lead to divergence.
		In the following, we suggest how to circumvent this problem, i.e., how to use the superposition principle without the need of a nomographic representation of the max-function.

	\section{Asymptotic Max-Consensus Protocol}
	\label{sec:asympt}
	Computing the average function over {an ideal MAC of form} (\ref{eq:MAC_ideal}) by exploiting the interference is straightforward, since the function is trivially nomographic (\cite{molinari2018exploiting}). In the following, we present a max-consensus protocol which makes use of the average function to achieve an agreement.
	The strategy is inspired by the following observation.
	\begin{proposition}  
		\label{core_idea_expl}
		Given a set of agents $\mathcal{N}$ and a non-empty subset $\mathcal{M}\subseteq\mathcal{N}$,
		$\forall k\in\mathbb{N}$, $\forall i\in\mathcal{N}$,
		$$x_i(k)<\frac{\sum_{j\in \mathcal{M}}x_j(k)}{|\mathcal{M}|} \implies x_i(k)<\max_{j\in\mathcal{N}}({x}_j(k)).$$
		\begin{proof}
			$\forall k\in\mathbb{N}$, $\forall i\in\mathcal{N}$,
			$x_i(k)<\frac{\sum_{j\in \mathcal{M}}x_j(k)}{|\mathcal{M}|} \implies \exists j\in \mathcal{M},\ j\not=i: x_j(k)>x_i(k)$, 
			which immediately implies $x_i(k)<\max_{j\in\mathcal{M}}({x}_j(k))\leq\max_{j\in\mathcal{N}}({x}_j(k))$.			
			\qed
		\end{proof}
	\end{proposition} 		
	
	Clearly, for max-consensus, any non-maximal agent does not need to broadcast its information state.
	In general, however, agents do not know whether they are maximal.
	We therefore settle for necessary conditions that can be locally evaluated.
	The result of this local evaluation for agent $i$ at time $k$ is stored in the boolean variable $\tilde{y}_i(k)$.
	If this variable is $1$, this means that agent $i$ satisfies the respective necessary condition at time $k$
	and it is said to be a \textbf{maximal candidate} at time $k$. 
	If (and only if) this is true, the agent will be allowed to broadcast at the next time instant.
	This will be expressed by an \textit{authorization variable} $y_i$, where $y_i(k)=\tilde{y}_i(k-1)$.

	Let
	$
		\forall i\in\mathcal{N},\ N_i^m(k) = \{j\in N_i\mid y_j(k)=1\}\subseteq N_i
	$
	be the set of neighboring agents of $i$ authorized to broadcast at time $k$.
	If $|N_i^m(k)|>0$, by Proposition \ref{core_idea_expl}, at time-step $k\in\mathbb{N}$, each agent $i\in\mathcal{N}$ for which
	\begin{equation}
		\label{eq:isMax}
		x_i(k)<\frac{\sum_{j\in N_i^m(k)}x_j(k)}{|N_i^m(k)|},
	\end{equation}
	cannot be a maximal candidate at time-step $k\in\mathbb{N}$.
	{Therefore $y_i$ can be updated as follows:} 
	\begin{multline}
		\forall i \in\mathcal{N},\ 
		 y_i(k+1)=\tilde{y}_i(k)=\\
		 =I_{\mathbb{R}_{\geq0}}\left( x_i(k)-\frac{\sum_{j\in N_i^m(k)}x_j(k)}{|N_i^m(k)|} \right).
	\end{multline}
	
	
	\subsection{Protocol Design}
	\label{subsec:comm_prot}	
	Under the assumption of a noiseless and non-fading channel, the following communication protocol, which makes use of an ideal MAC of order 2 (see \cite{goldenbaum2013harnessing}), is employed.
	At every time-step $k\in\mathbb{N}$, each agent $j\in\mathcal{N}$ broadcasts two orthogonal signals, $\tau_j(k)=y_j(k)x_j(k)$ and $\tau_j'(k)=y_j(k)$.
	
	Each agent $i\in\mathcal{N}$ receives two mutually orthogonal signals from its neighbors, 
	\begin{align}
		\label{eq:zeta}
		&z_i(k) = \sum_{j\in N_i}y_j(k)x_j(k)=\sum_{j\in N_i^m(k)}x_j(k)\\
		\text{and}\nonumber&\\
		&z_i'(k) = \sum_{j\in N_i}y_j(k)=\sum_{j\in N_i^m(k)}1=|N_i^m(k)|
		\label{eq:zetaprime}
	.
	\end{align}
	By making use of these two signals, each agent $i\in\mathcal{N}$, at every time $k\in\mathbb{N}$, can compute the average of information states of agents in $N_i^m(k)$, i.e.
	\begin{align}
		\label{eq:u_signal}
		u_i(k)=
		\begin{cases}			
			\frac{z_i(k)}{z_i'(k)}=\frac{\sum_{j\in N_i^m(k)}x_j(k)}{|N_i^m(k)|} &\text{if }N_i^m(k)\not=\emptyset\\
			0 &\text{else}.
		\end{cases}
	\end{align}
	Notice that interference has been exploited for computing $u_i(k)$.
	
	
	\subsection{Controller Design}	
	Each agent $i\in\mathcal{N}$ is endowed with the following consensus dynamics:
	\begin{equation}
		\label{eq:asympt_cons_prot}
		\forall k\in\mathbb{N}:\ \ \ 
		\begin{cases}
			x_i(k+1)=\max(x_i(k), u_i(k))\\
			y_i(k+1)=I_{\mathbb{R}_{\geq0}}(x_i(k)-u_i(k))
		\end{cases},
	\end{equation}
	where $y_i(1)=1$, $x_i(1)=x_{i_0}$, {and $u_i(k)$ computed as in} (\ref{eq:u_signal}). 	
	Note that $u_i(k)$ depends on both $x_i(k)$ and $y_i(k)$.
	In vector-form, (\ref{eq:u_signal})-(\ref{eq:asympt_cons_prot}) become
	\begin{equation}
		\label{eq:asympt_cons_prot_matForm}
		\mathbf{w}(k+1)=
		g(\mathbf{w}(k))
		,
	\end{equation}
	where
	\begin{equation}
		\mathbf{w}(k)=
		\begin{bmatrix}
			\mathbf{x}(k)\\
			\mathbf{y}(k)
		\end{bmatrix},
	\end{equation}
	and, $\forall i\in\mathcal{N},\ [\mathbf{y}(k)]_i=y_i(k)$ and $g:\mathbb{R}^{n}\times\{0,1\}^n\rightarrow\mathbb{R}^{n}\times\{0,1\}^n$ the corresponding nonlinear map. 
	{Using Lyapunov theory}, e.g. \cite[p. 87]{aastrom2013computer}, the convergence of (\ref{eq:asympt_cons_prot_matForm}) to max-consensus can be formalized. 
	To this end, we need to study some properties of the system.
	
	\begin{proposition} 
		\label{prop:xincreas}
		Given a network topology $(\mathcal{N},\mathcal{A})$ and 
		a consensus dynamics (\ref{eq:asympt_cons_prot_matForm}),
		$			
			\forall
			\mathbf{x}(1)
			\in
			\mathbb{R}_{\geq0}^{n}
			,\ \forall i\in\mathcal{N},\ 
			\forall k\in\mathbb{N},\ 
			x_i(k)\leq x_i(k+1)\leq \max(\mathbf{x}(1)).			
		$
		
		\begin{proof}
			By (\ref{eq:u_signal}), $\forall i\in\mathcal{N}$, $\forall k\in\mathbb{N}$, $u_i(k)\leq\max(\mathbf{x}(k))$.
			Hence with (\ref{eq:asympt_cons_prot}), 
			the iteration (\ref{eq:asympt_cons_prot_matForm}) {generates a non-decreasing bounded sequence}
			$x_i(0)\leq x_i(1)\leq\dots\leq x_i(k)\leq x_i(k+1) \leq \max(\mathbf{x}(k+1))$, $1\leq i\leq n$.
			This implies $x_i(k)\leq x_i(k+1)\leq \max(\mathbf{x}(k+1))=\max(\mathbf{x}(1))$.
			\qed
		\end{proof}		
	\end{proposition} 
	\begin{proposition} 
		\label{prop:equil}
		If $(\mathcal{N},\mathcal{A})$ is a connected graph, 
		$
		\mathbf{w}^*=
		\begin{bmatrix}
			\mathbf{x}^*\\
			\mathbf{1}_n
		\end{bmatrix}
		$,
		where
		$
		\mathbf{x}^*
		=
		x^*\mathbf{1}_n
		,\
		x^*=\max(\mathbf{x}(0))
		,
		$
		is an equilibrium point for (\ref{eq:asympt_cons_prot_matForm}),
		
		
		\begin{proof}
			{Clearly}, 
			$
			\mathbf{w}^*=
			\begin{bmatrix}
				{\mathbf{x}^*}',
				{\mathbf{y}^*}'
			\end{bmatrix}'
			$
			is an equilibrium point if and only if, $\forall i\in\mathcal{N}$,
			\begin{align}
				\label{eq:1_proof_eq}
				\mathbf{x}_i^*&=\max(\mathbf{x}_i^*,\mathbf{u}_i^*)\\
				\label{eq:2_proof_eq}
				\mathbf{y}_i^*&=I_{\mathbb{R}_{\geq0}}(\mathbf{x}_i^*-\mathbf{u}_i^*)
			\end{align}
			where 
			\begin{align}
				\mathbf{u}_i^*=
				\begin{cases}
					\frac{\sum_{j\in N_i^m}\mathbf{x}_j^*}{|N_i^m|}	&\text{if }|N_i^m|\not=0\\
					0										&\text{else.}
				\end{cases}
			\end{align}
			For $\mathbf{x}^*=x^*\mathbf{1}_n$ and $\mathbf{y}^*=\mathbf{1}_n$, $N_i^m=N_i$.
			As $(\mathcal{N},\mathcal{A})$ is connected, $|N_i|\geq1$.
			Therefore, $\mathbf{u}_i^*={x}^*$ and (\ref{eq:1_proof_eq})-(\ref{eq:2_proof_eq}) are satisfied.
			
			\qed
		\end{proof}
	\end{proposition} 
	\begin{corollary}
		\label{prop:equil_unique}
		Given a connected network topology $(\mathcal{N},\mathcal{A})$, then 
		$
		\forall
		\mathbf{x}(1)
		\in
		\mathbb{R}_{\geq0}^{n}
		$
		, 
		$
		\mathbf{w}^*
		$ 
		is the unique equilibrium point of (\ref{eq:asympt_cons_prot_matForm}).
		
		\begin{proof}
			The proof is by contradiction. 
			Hence, assume that $\mathbf{w}^*=[{\mathbf{x}^*}',{\mathbf{y}^*}']'$
			with $\mathbf{x^*}=x^*\mathbf{1}_n$
			and ${\mathbf{y}^*}\not=\mathbf{1}_n$ is an equilibrium point. 
			The latter implies that $\mathbf{y}_i^*=0$ for at least one $i\in\mathcal{N}$.
			Hence, from (\ref{eq:2_proof_eq}),
			$\mathbf{x}_i^*<\mathbf{u}_i^*$
			and (\ref{eq:1_proof_eq}) is violated.
			This establish that $\mathbf{w}^*$ with ${\mathbf{y}^*}\not=\mathbf{1}_n$ is not an equilibrium point. 			
			Now consider the case $\mathbf{x^*}\not=x^*\mathbf{1}_n$ and ${\mathbf{y}^*}=\mathbf{1}_n$.
			Choose $i$ such that $\mathbf{x}_i^*=\min(\mathbf{x}^*)$.
			As $\mathbf{x^*}\not=x^*\mathbf{1}_n$, $\mathbf{x}_i^*<x^*$.
			On the other hand, by Proposition \ref{prop:xincreas}, an agent that is maximal at $k=1$
			remains so $\forall k\in\mathbb{N}$. 
			Hence, there exists $l\in\mathcal{N}$ such that $\mathbf{x}_l^*=x^*>\mathbf{x}_i^*$.
			Then, as $(\mathcal{N},\mathcal{A})$ is connected, there exists $(i,j)\in\mathcal{A}$ such that $\mathbf{x}_j^*>\mathbf{x}_i^*$.
			As ${\mathbf{y}}^*=\mathbf{1}_n$, $N_i^m=N_i$ and therefore $\mathbf{u}_i^*>\mathbf{x}_i^*$. 
			This violates (\ref{eq:1_proof_eq}) and therefore contradicts the assumption. 
			\qed
		\end{proof}
	\end{corollary}
	
	
	\begin{proposition} 
		\label{prop:contraction}
		A connected network topology $(\mathcal{N},\mathcal{A})$ is given. Then, 
		$
		\forall
		\mathbf{x}(1)
		\in
		\mathbb{R}_{\geq0}^{n}
		$
		, 
		$\forall k\in\mathbb{N}$,
		\begin{equation}
			\label{eq:prop_diff2_0}
			\sum_{i\in\mathcal{N}} \left( x_i(k+2)-x_i(k) \right) =0
			\implies 
			\mathbf{x}(k)=\mathbf{x}^*
			.
		\end{equation}
		
		\begin{proof}
				By Proposition \ref{prop:xincreas}, $\{ x_i(k) \}_{k\in\mathbb{N}} $ is a non-decreasing sequence of nonnegative entries. 
				Therefore, $\sum_{i\in\mathcal{N}} (x_i(k+2)-x_i(k))=0$ if and only if $\mathbf{x}(k)=\mathbf{x}(k+1)=\mathbf{x}(k+2)$, 
				which implies (because of (\ref{eq:asympt_cons_prot})), $\forall i\in\mathcal{N},\ x_i(k)\geq u_i(k)$ and $x_i(k+1)\geq u_i(k+1)$ and therefore $\mathbf{y}(k+1)=\mathbf{y}(k+2)=\mathbf{1}_n$. 
				Therefore, $\mathbf{w}(k+1)=\mathbf{w}(k+2)$, which is possible if and only if $\mathbf{w}(k+1)=\mathbf{w}^*$ due to Corollary \ref{prop:equil_unique}.
				$\mathbf{w}(k+1)=\mathbf{w}^*$ is equivalent to $\mathbf{x}(k+1)=\mathbf{x}^*$ and $\mathbf{y}(k+1)=\mathbf{1}_n$. 
				The latter implies that $\mathbf{x}(k)=\mathbf{u}(k)$ and therefore $\mathbf{x}(k)=\mathbf{x}(k+1)=\mathbf{x}^*$.
				\qed
		\end{proof}		
	\end{proposition} 
	By Proposition \ref{prop:contraction} and Proposition \ref{prop:xincreas}, 
	$$\mathbf{x}(k)\not=\mathbf{x}^*\implies \sum_{i\in\mathcal{N}}( x_i(k+2)-x_i(k))>0.$$

\begin{figure}[t]
		\centering
		\begin{subfigure}{.49\columnwidth}
			\centering
			\includegraphics[width=\columnwidth]{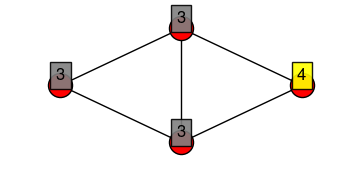}
			\caption{$(\tilde{\mathcal{N}},\tilde{\mathcal{A}})$ and $\mathbf{x}_{(1)}(0)$.}
			\label{fig:graph_as_conv}
		\end{subfigure}
		\begin{subfigure}{.49\columnwidth}
			\centering
			\includegraphics[width=\columnwidth]{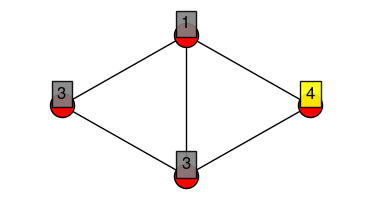}
			\caption{$(\tilde{\mathcal{N}},\tilde{\mathcal{A}})$ and $\mathbf{x}_{(2)}(0)$.}
			\label{fig:graph_as_conv_diffx0}
		\end{subfigure}
		\caption{Topology and initial conditions for the examples.}
	\end{figure}

\begin{figure}[h]
		\centering
		\begin{subfigure}{\columnwidth}
			\includegraphics[width=\columnwidth]{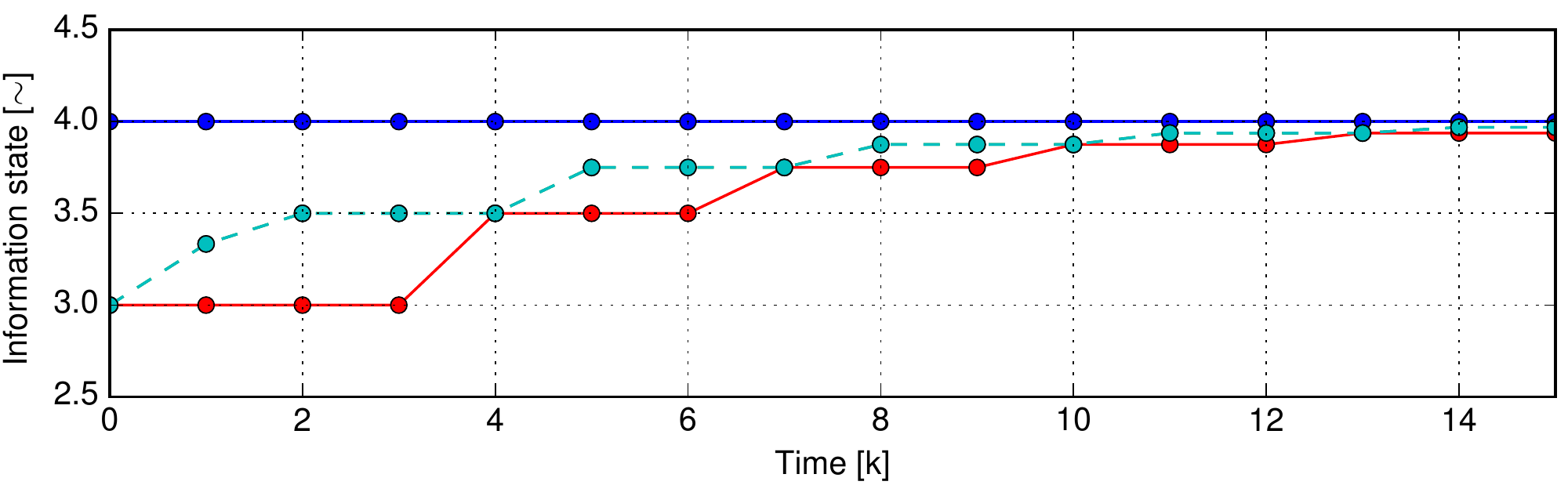}
			\caption{Asymptotic convergence of $x_i(k)$.}
			\label{fig:as_conv_plot}
		\end{subfigure}
		\begin{subfigure}{\columnwidth}
			\includegraphics[width=\columnwidth]{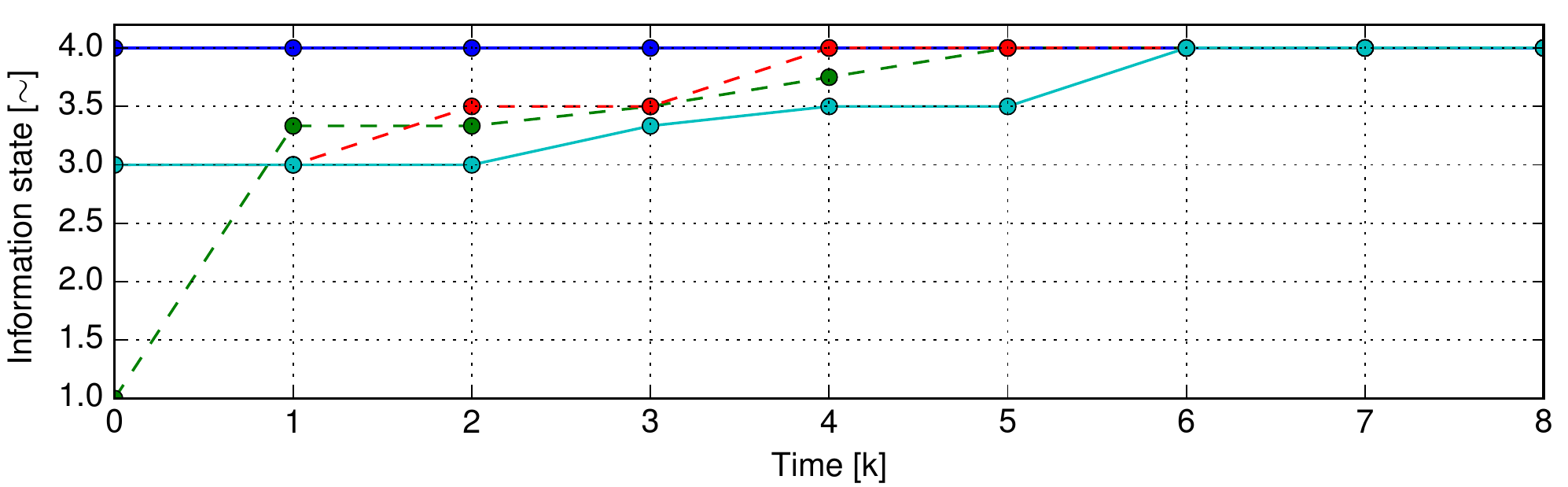}
			\caption{Finite-time agreement in $6$ steps.}
			\label{fig:as_conv_plot_diffx0}
		\end{subfigure}
		\caption{Evolution of information states for the Examples \ref{example:asympt} (top) and \ref{example:finite-time} (bottom).}
	\end{figure}

	\begin{proposition} 
		\label{prop:lyap}
		Given a connected network topology $(\mathcal{N},\mathcal{A})$. For every initial state $\mathbf{x}(1)\in\mathbb{R}_{\geq0}$, 
		the consensus protocol (\ref{eq:u_signal})-(\ref{eq:asympt_cons_prot}) converges asymptotically to max-consensus.
		
		\begin{proof}			
			By (\ref{eq:asympt_cons_prot}), 
			$ y_i(k+1)=1 \iff x_i(k)\geq u_i(k)$
			and
			$ x_i(k+1)=x_i(k) \iff x_i(k)\geq u_i(k).$
			Therefore, $y_i(k+1)\iff(x_i(k+1)=x_i(k))$, which can lead to rewriting $y_i(k)$ as
			$y_i(k)=I_{\mathbb{R}_{\geq0}}(x_i(k-1)-x_i(k))$.			
			Hence, we can rewrite (\ref{eq:asympt_cons_prot_matForm})
			such that the system dynamics is described by the state 
			${\mathbf{v}}(k)$
			given by
			$
			{\mathbf{v}}(k)
			=
			\left[
			\mathbf{x}(k)',\ 
			\mathbf{x}(k-1)'
			\right]',
			$
			and by {the nonlinear map $\tilde{g}$}, 
			\begin{equation}
			\label{eq:equil_state_change}
			{\mathbf{v}}(k+1)
			=
			\tilde{g}(			
			{\mathbf{v}}(k)
			).
			\end{equation}
			The equilibrium $\mathbf{w}^*$ of (\ref{eq:asympt_cons_prot_matForm}) corresponds in (\ref{eq:equil_state_change}) to the equilibrium
			$\mathbf{v}^*
			=
			\left[
				\mathbf{x}^{*'},\ 
				\mathbf{x}^{*'}
			\right]'
			=
			x^*\mathbf{1}_{2n}.
			$		
			The following analysis is based on \cite[p. 88]{aastrom2013computer} and \cite[p. 22]{lalo2014advanced}, which formalize the Lyapunov method for discrete-time systems.
			Let
			\begin{align*}
				V(\mathbf{v(k)})&=2n\max(\mathbf{v}(k))-\mathbf{1}_{2n}'\mathbf{v}(k)\\
				&=2nx^* - \sum_{i\in\mathcal{N}}x_i(k)+x_i(k-1)\numberthis
			\end{align*}
			be a candidate Lyapunov function for (\ref{eq:equil_state_change}).
			The following properties hold:
			\begin{itemize}
				\item[a)] $V(\mathbf{v})$ is continuous in $\mathbb{R}_{\geq0}^{2n}$;
				\item[b)] $V(\mathbf{v}^*)=0$;
				\item[c)] By Proposition \ref{prop:xincreas}, 
				$V(\mathbf{v})$ is positive definite on any trajectory of $\mathbf{v}$ satisfying (\ref{eq:equil_state_change}) unless $\mathbf{v}=\mathbf{v}^*$;
				\item[d)] By Proposition \ref{prop:contraction}, $\forall \mathbf{v}(k)\not=\mathbf{v}^*$,
					\begin{align*}
						\Delta V(\mathbf{v}(k))&=V(\mathbf{v}(k+1))-V(\mathbf{v}(k))\\
						&=\sum_{i\in\mathcal{N}}(x_i(k-1)-x_i(k+1))<0.
					\end{align*}
			\end{itemize}
			Since $V(\mathbf{v}(k))$ is a generalized energy function for (\ref{eq:equil_state_change})
			and it strictly decreases along the trajectories of the system, 
			by \cite[p. 87]{aastrom2013computer}, 
			the state $\mathbf{v}(k)$ will asymptotically converge to $\mathbf{v}^*$. 
			\qed
		\end{proof}
	\end{proposition} 
	\begin{corollary}
		\label{prop:soonerOrLater}
		Given a connected network topology $(\mathcal{N},\mathcal{A})$, the following holds
		\begin{multline}
			\forall
			\mathbf{x}(1)
			\in
			\mathbb{R}_{\geq0}^{n},\ 
			\forall i\in\mathcal{N},
			\\
			x_i(1)<\max(\mathbf{x}(1))
			\implies
			\exists k_i\in\mathbb{N}:\ y_i(k_i)=0.			
		\end{multline}
		
		\begin{proof}
			By Proposition \ref{prop:lyap}, $\forall i\in\mathcal{N}$,
			$$
				\lim\limits_{k\rightarrow\infty}x_i(k)=x^*=\max(\mathbf{x}(1)).
			$$
			As a consequence, $\forall i\in\mathcal{N}$, 
			$$
			x_i(1)<\max(\mathbf{x}(1))
				\implies
			\exists k_i\in\mathbb{N}:\ x_i(k_i)>x_i(k_i-1),
			$$
			which implies $y_i(k_i)=0$ (see proof of Proposition \ref{prop:lyap}).
			\qed
		\end{proof}
	\end{corollary}
	\begin{example}
		\label{example:asympt}
		Let a multi-agent system be modeled by a connected communication topology $(\tilde{\mathcal{N}},\tilde{\mathcal{A}})$ which is represented in Figure \ref{fig:graph_as_conv}. The vector of initial information states is $\mathbf{x}_{(1)}(1)=[4,3,3,3]'$. A simulation of (\ref{eq:u_signal})-(\ref{eq:asympt_cons_prot}) is shown in Figure \ref{fig:as_conv_plot}.
	\end{example}	
	
	We have shown that the suggested consensus protocol achieves asymptotic convergence to the max consensus
	for arbitrary initial information states if the network topology is represented by a connected graph. 
	Moreover, extensively numerical experiments have indeed verified that in most cases finite time convergence is achieved. 
	This is demonstrated in the following example. 
	\begin{example}
		\label{example:finite-time}
		A multi-agent system with communication topology $(\tilde{\mathcal{N}},\tilde{\mathcal{A}})$, as in Example \ref{example:asympt}, is given. However, the vector of initial information states is now $\mathbf{x}_{(2)}(1)=[4,3,1,3]'$.
		The system shows finite-time convergence, as shown in Figure \ref{fig:as_conv_plot_diffx0}.
	\end{example}		
	
	\section{Finite-Time Max-Consensus Protocol }
	\label{section:finite}
	By using orthogonal channel access methods and the standard max-consensus protocol (\ref{eq:maxcons}), 
	max-consensus is always achieved in a finite number of steps if the underlying graph is connected (\cite{nejad2009max}).
	With the consensus protocol (\ref{eq:asympt_cons_prot}), max-consensus is achieved by exploiting interference.
	However, in general, consensus will be reached asymptotically.
	
	Based on the consensus protocol in Section \ref{subsec:comm_prot}, 
	we can suggest a switching consensus dynamics that will exploit the superposition property of the wireless channel
	and achieve finite-time max-consensus for arbitrary vectors of initial states. 	
	\begin{subequations}\label{eq:switching_top_dyn}
		\begin{align*}
			\intertext{\textbullet\ if $k = 2T_i(k):$}
			&\begin{cases}
				x_i(k+1)=\max(x_i(k), u_i(k))\\
				y_i(k+1)=\Pi_{t=T_i(k)}^{k}y_i(t)\\
				T_i(k+1)=k
			\end{cases} 
			\label{eq:switching_top_dyn1} \numberthis&,
			\intertext{\textbullet\ else:}
			&\begin{cases}
				x_i(k+1)=\max(x_i(k), u_i(k))\\
				y_i(k+1)=I_{\mathbb{R}_{\geq0}}(x_i(k)-u_i(k))\\
				T_i(k+1)=T_i(k)
			\end{cases} 
			\label{eq:switching_top_dyn2} \numberthis&.
		\end{align*}
	\end{subequations}
	An additional state variable, $T_i:\mathbb{N}\rightarrow\mathbb{N}$, has been added to the system and has initial conditions $T_i(1)=1$.
	The initial conditions $x_i(1)$ and $y_i(1)$ are the same as in (\ref{eq:asympt_cons_prot}).
	
	The controller switches between two dynamics: it keeps (\ref{eq:switching_top_dyn2})
	(which has the same dynamics as (\ref{eq:asympt_cons_prot})) in every time step with the exception of $k=2^{p},\ p\in\mathbb{N}$,
	when the protocol has dynamics (\ref{eq:switching_top_dyn1}).
	In this case, $T_i$ is updated by $T_i(k+1)=2T_i(k)$. 
	Moreover, $y_i(k+1)$ is computed so that any agent that has not been a
	\textit{maximal candidate} in all of its last $T_i(k)$ time steps
	will not be a maximal candidate at time $k+1$.

%
	\begin{proposition} 
		\label{prop:finite_time}		
		Let $(\mathcal{N},\mathcal{A})$ be a connected graph.
		Then, the switching consensus protocol (\ref{eq:switching_top_dyn}) achieves finite-time max-consensus
		for any $\mathbf{x}(1)\in\mathbb{R}_{\geq0}$.
		
		\begin{proof}
			System (\ref{eq:switching_top_dyn}) has the same dynamics as (\ref{eq:asympt_cons_prot}) in all time steps except for $k=2^{p},\ p\in\mathbb{N}$, in which the system dynamics is (\ref{eq:switching_top_dyn1}).
			If, at time $k\in\mathbb{N}$, all non-maximal agents are not 
			authorized to broadcast
			(i.e. $\forall i\in\mathcal{N}, x_i(k)<x^*\implies y_i(k)=0$), 
			all agents in set
			$$
			\mathcal{L}_1(k)=\{ i\in N_j\mid  x_j(k)=x^* \}
			$$
			will get to max-consensus at step $k+1$. 
			By Corollary \ref{prop:soonerOrLater},
			$\exists \tilde{k}\in\mathbb{N}$ large enough, such that, $\forall i\in\mathcal{N}$,
			\begin{multline*}
				x_i(T_i(\tilde{k}))<x^*
					\implies
				\exists k_i\in[T_i(\tilde{k}),\tilde{k}]:\ y_i(k_i)=0.
			\end{multline*}
			As a consequence and according to (\ref{eq:switching_top_dyn1}), 
			$\forall i\in\mathcal{N}$, 
			$$
				x_i(T_i(\tilde{k}))<x^*
				\implies
				y_i(2T_i(\tilde{k})+1)=0,
			$$
			which implies that 
			$$
				\forall i\in\mathcal{L}_1(2T_i(\tilde{k})),\ x_i(2T_i(\tilde{k})+2)=x^*.
			$$
			As $|\mathcal{N}|$ is finite, within a finite number of {recursions}, max-consensus is achieved.
			\qed
		\end{proof}
	\end{proposition} 
	\section{Simulation}
	\label{sec:simul}
	
	\subsection{Scenario}
\begin{figure}[h]
		\centering
		\includegraphics[width=\columnwidth]{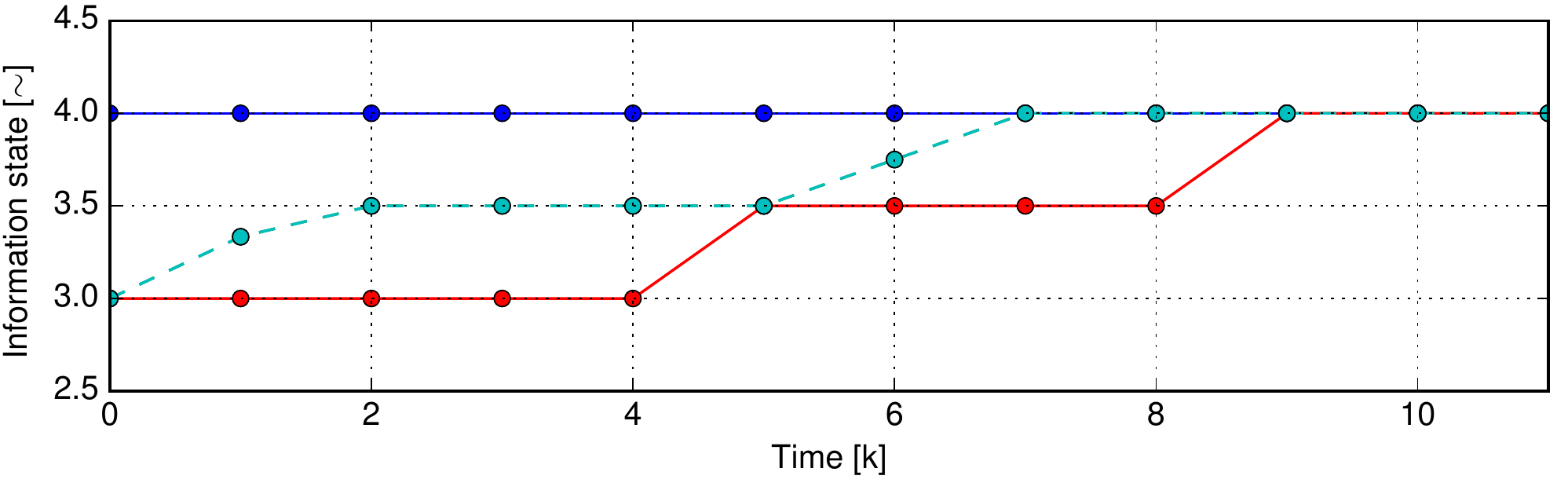}
		\caption{System in Figure \ref{fig:graph_as_conv} endowed with (\ref{eq:switching_top_dyn}) achieves finite-time agreement.}
		\label{fig:fin_conv_plot}
	\end{figure}

	The multi-agent system with communication graph $(\tilde{\mathcal{N}},\tilde{\mathcal{A}})$ and with initial state $\mathbf{x}_{(1)}(1)$ as
	in Figure \ref{fig:graph_as_conv} converges asymptotically to the agreement.
	As claimed in Section \ref{section:finite}, by using (\ref{eq:switching_top_dyn}),
	any system with a connected communication topology
	achieves finite-time consensus.
	The agents in $(\tilde{\mathcal{N}},\tilde{\mathcal{A}})$ are therefore endowed with the switching consensus dynamics (\ref{eq:switching_top_dyn}).
	In Figure \ref{fig:fin_conv_plot}, the agents' information states are plotted
	and shown to achieve agreement in $\bar{k}=9$ steps.\
	
	\subsection{Randomized Scenarios}
\begin{figure}
		\centering
		\begin{subfigure}{\columnwidth}
			\centering
			\includegraphics[width=\columnwidth]{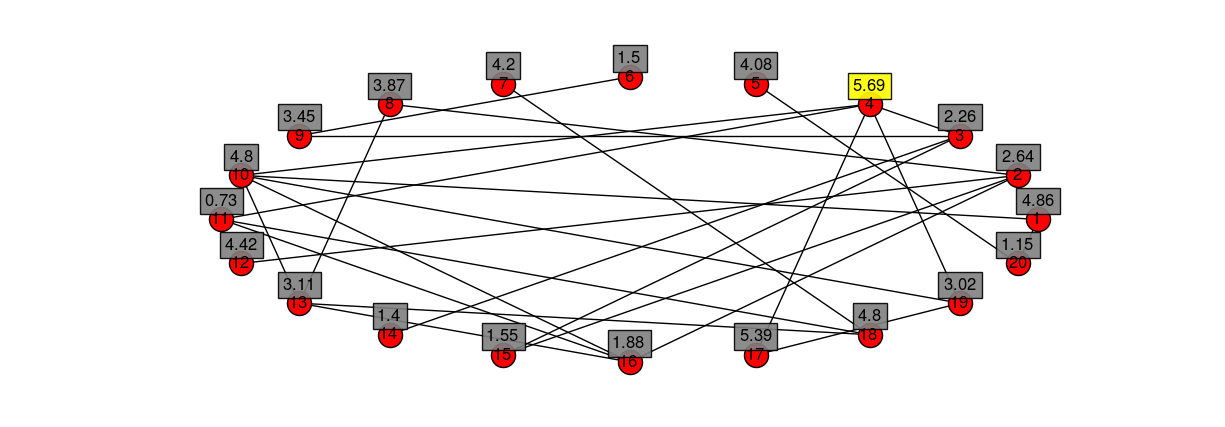}
			\caption{$({\mathcal{N}}_l,{\mathcal{A}}_l)$ and $\mathbf{x}_{l}(0)$.}
			\label{fig:random_graph}
		\end{subfigure}
		\begin{subfigure}{\columnwidth}
			\includegraphics[width=\columnwidth]{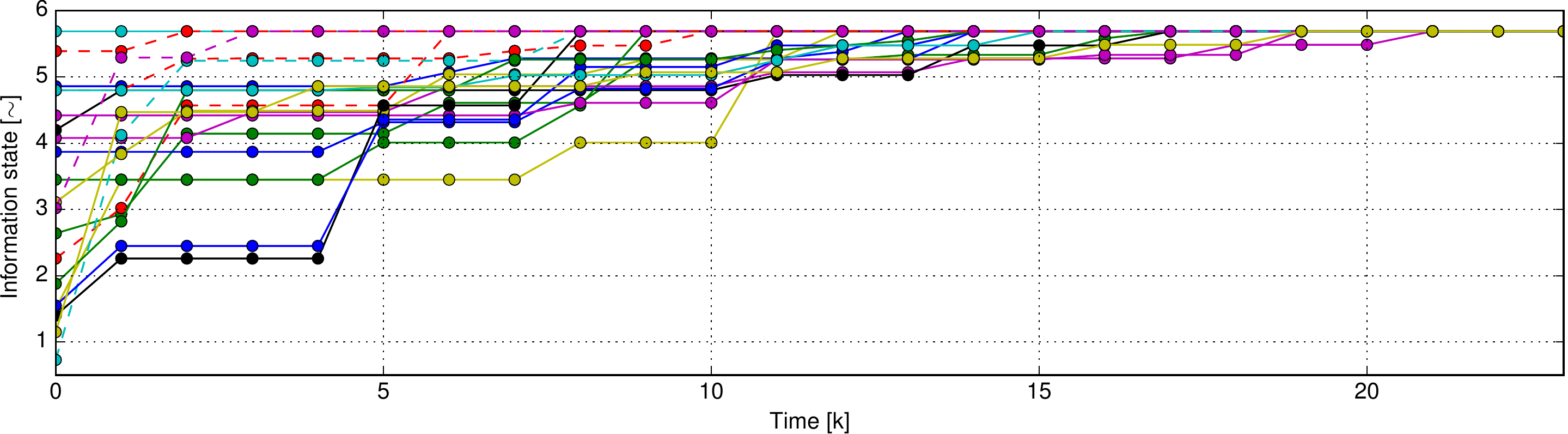}
			\caption{Finite-time convergence of $x_i(k)$.}
			\label{fig:random_plot}
		\end{subfigure}
		\caption{System with dynamics (\ref{eq:switching_top_dyn}).}
	\end{figure}

	A multi-agent system endowed with consensus protocol (\ref{eq:switching_top_dyn}) and 
	with network topology $(\mathcal{N}_l,\mathcal{A}_l)$, with initial state vector $\mathbf{x}_l(1)$,
	as in Figure \ref{fig:random_graph},
	is simulated.
	The network size is $|\mathcal{N}_l|=20$, and its communication topology is connected.
	Each entry of the initial information state vector $\mathbf{x}_l(1)$ is drawn from a uniform distribution $\mathcal{U}(0,2\pi)$.
	Figure \ref{fig:random_plot} shows that agents get to consensus in a finite number of steps, i.e.
	$\forall i\in\mathcal{N}_l,\ x_{i}(\bar{k})=\max(\mathbf{x}_l(0))$. In this example, $\bar{k}=21$.	

	\subsection{Comparison with traditional solution}
	\begin{figure}[h]
		\includegraphics[width=\columnwidth]{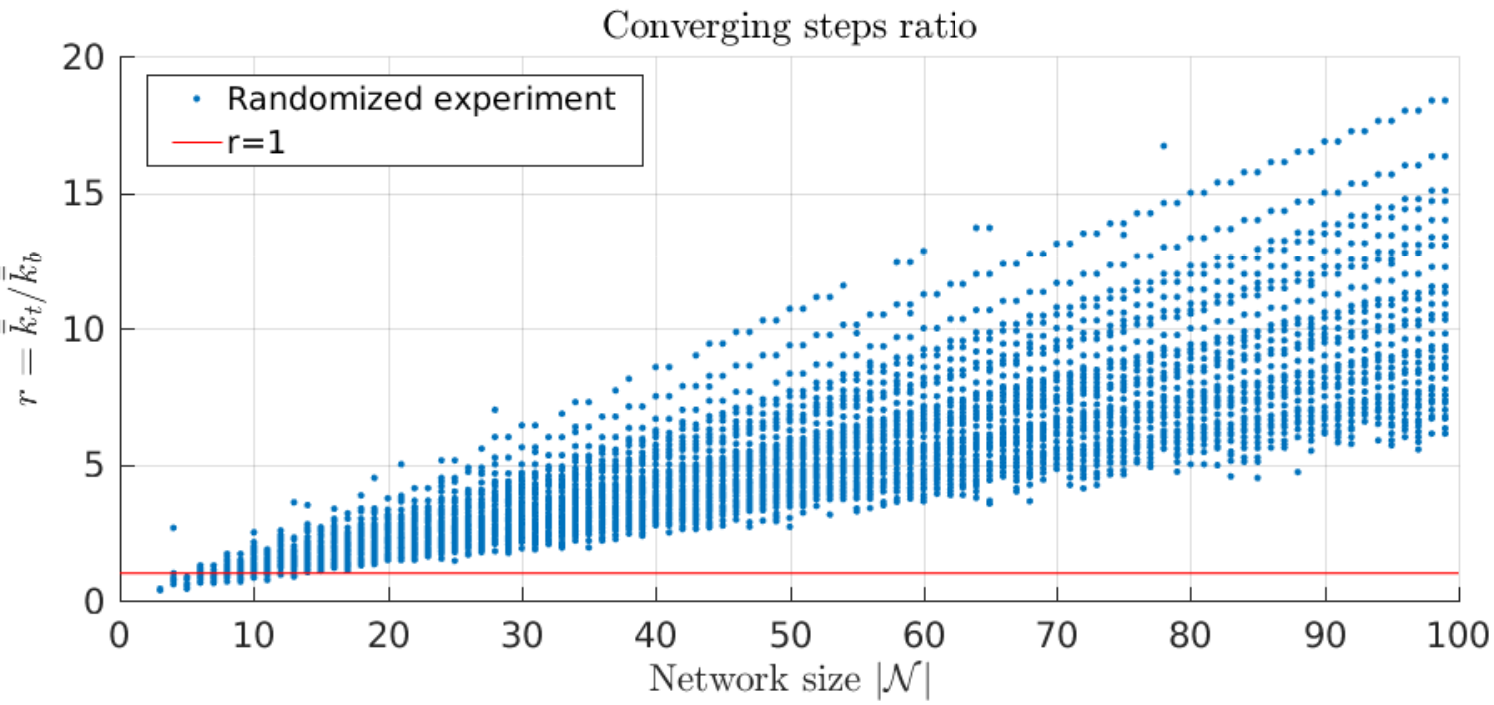}
		\caption{Ratio of steps required for converging with traditional ($\bar{k}_{t}$) and broadcast protocols ($\bar{k}_{b}$). }
		\label{fig:_TDMA}
	\end{figure}
	In the following, by \textit{traditional protocols}, we denote 
	those strategies which use orthogonal channel access methods such as TDMA (Time Division Multiple Access) for communication and a consensus dynamics like (\ref{eq:maxcons}).
	With TDMA, the transmission is multiplexed in time-domain. 
	Each agent will then be allowed to transmit its information state in a pre-assigned time slot (this will avoid interference between signals coming from different agents).
	Accordingly, for a network modeled by topology $(\mathcal{N},\mathcal{A})$, $|\mathcal{N}|$ time slots should be assigned one-to-one to all nodes within one sampling interval.
	Intuitively, each run of (\ref{eq:maxcons})
	corresponds to $\frac{|\mathcal{N}|}{2}$ runs of (\ref{eq:switching_top_dyn}),
	since the latter uses a communication protocol based on the broadcast of only two orthogonal signals,
	as presented in Section \ref{subsec:comm_prot}.
	A comparison (via randomized analysis) between traditional protocols and the one proposed in this paper in terms of steps required for converging
	is presented in Figure \ref{fig:_TDMA}.
	
	Given a topology and an initial state vector, $\bar{k}_{t}$ denotes the number of steps required to achieve consensus with traditional protocols; 
	$\bar{k}_{b}$ represents, for the same problem, the number of steps required to achieve consensus with broadcast protocol (\ref{eq:switching_top_dyn}).
	Their ratio is $r=\frac{\bar{k}_{t}}{\bar{k}_{b}}$.
	For networks with more than circa $20$ agents, adopting broadcast solutions gives faster convergence. 
	In case $|\mathcal{N}|=100$, broadcast algorithm (\ref{eq:switching_top_dyn}) achieves consensus between $5$ and $20$ times faster than traditional approaches.
	
	\section{Conclusion}
	\label{sec:concl}
	In this paper, 
	a consensus protocol for reaching max-consensus in a finite number of steps
	is suggested.
	Its main characteristic is that it employs the broadcast and superposition properties of the wireless channel.
	This can drastically reduce the number of required messages. 
	Indeed
	simulations show that this algorithm exhibits considerably faster convergence
	than traditional approaches.
	The wireless channel has been modeled by an ideal MAC.
	Future work will analyze the impact of nonidealities, such as channel coefficients and receiver noise.
	
\bibliography{biblio}
\bibliographystyle{IEEEtran}
\end{document}